\documentclass[conference]{IEEEtran}
\IEEEoverridecommandlockouts

\usepackage{cite}
\usepackage{amsmath,amssymb,amsfonts}
\usepackage{algorithmic}
\usepackage{graphicx}
\usepackage{textcomp}
\usepackage{xcolor}
\usepackage{hyperref}       
\usepackage{url}            
\usepackage{booktabs}       
\usepackage{nicefrac}       
\usepackage{microtype}      
\usepackage{lipsum}
\usepackage{caption}
\usepackage{subcaption}
\usepackage{makecell}
\usepackage{multirow}

\graphicspath{ {./} }

\def\BibTeX{{\rm B\kern-.05em{\sc i\kern-.025em b}\kern-.08em
    T\kern-.1667em\lower.7ex\hbox{E}\kern-.125emX}}
\begin{document}

\title{
    FlashRecovery: Fast and Low-Cost Recovery from Failures for Large-Scale Training of LLMs
}

\author{
    \IEEEauthorblockN{
        Haijun Zhang$^{1,2}$, Jinxiang Wang$^1$, Zhenhua Yu$^1$, Yanyong Zhang$^2$, Xuejie Ji$^1$, Kaining Mao$^1$, \\ 
        Jun Zhang$^1$, Yaqing Zhang$^1$, Ting Wu$^{1,2}$, Fei Jie$^1$, Xiemin Huang$^1$, Zhifang Cai$^3$, \\
        Junhua Cheng$^3$, Shuwei Wang$^3$, Wei Li$^3$, Xiaoming Bao$^3$, Hua Xu$^3$, Shixiong Zhao$^3$, \\
        Jun Li$^3$, Hongwei Sun$^3$, Ziyang Zhang$^3$, Yi Xiong$^3$, Chunsheng Li$^3$
    }
    \IEEEauthorblockA{
      \textit{$^1$iFLYTEK AI Engineering Institute}, Hefei 230088, China \\
      \textit{$^2$University of Science and Technology of China}, Hefei 230026, China \\
        \textit{$^3$Huawei Technologies Co., Ltd}, Shenzhen 518129, China \\
    }
}

\maketitle

\begin{abstract}
    Large language models (LLMs) have made a profound impact across various fields due to their advanced capabilities.
    However, training these models at unprecedented scales requires extensive AI accelerator clusters and sophisticated parallelism strategies, which pose significant challenges in maintaining system reliability over prolonged training periods.
    A major concern is the substantial loss of training time caused by inevitable hardware and software failures.
    To address these challenges, we present FlashRecovery, a fast and low-cost failure recovery system comprising three core modules: 
    (1) Active and real-time failure detection. This module performs continuous training state monitoring, enabling immediate identification of hardware and software failures within seconds, thus ensuring rapid incident response;
    (2) Scale-independent task restart. By employing different recovery strategies for normal and faulty nodes, combined with an optimized communication group reconstruction protocol, our approach ensures that the recovery time remains nearly constant, regardless of cluster scale;
    (3) Checkpoint-free recovery within one step. Our novel recovery mechanism enables single-step restoration, completely eliminating dependence on traditional checkpointing methods and their associated overhead.
    Collectively, these innovations enable FlashRecovery to achieve optimal Recovery Time Objective (RTO) and Recovery Point Objective (RPO), substantially improving the reliability and efficiency of long-duration LLM training.
    Experimental results demonstrate that FlashRecovery system can achieve training restoration on training cluster with 4, 800 devices in 150 seconds. We also verify that the time required for failure recovery is nearly consistent for different scales of training tasks.
\end{abstract}


\section{Introduction}

Large language models (LLMs) have emerged as a focal point in artificial intelligence (AI) research due to their exceptional capabilities across a range of tasks and applications, including text generation, code assistance, and scientific discovery \cite{achiam2023gpt,liu2023your,song2023deepspeed4science}.

The superior performance of these models is fundamentally governed by scaling laws, with empirical studies consistently demonstrating that their capabilities improve as model parameters and training data increase \cite{kaplan2020scaling}. However, this growth in model scale necessitates the use of increasingly large clusters of GPUs or specialized AI accelerators, along with extended training duration. Consequently, maintaining reliable LLM training in large-scale accelerator clusters over extended periods presents a critical yet formidable challenge.

The increasing scale of LLMs has been enabled by advances in parallelism techniques, including data parallelism, model parallelism, pipeline parallelism, and hybrid three-dimensional parallelism \cite{rajbhandari2020zero,li2023colossal,shoeybi2019megatron}. 
While these methods facilitate large-scale model training, they also introduce significant vulnerability to errors due to their inherently synchronous nature. In such frameworks, even a single node failure can disrupt the entire training process, necessitating a full cluster halt and costly re-computation during recovery. For instance, in data parallelism, the training data is partitioned across clusters along the batch dimension, with each accelerator independently processing its assigned data segment through forward and backward passes. Then a synchronous collective communication is performed to aggregate gradients globally before updating model parameters. Similarly, pipeline parallelism divides the model into sequential stages, requiring intermediate tensor exchanges at predetermined pipeline boundaries, while tensor parallelism distributes layer parameters across multiple dimensions, necessitating continuous communication of activation tensors. Although these strategies enable efficient distributed training, their dependence on strict synchronization exacerbates the cost of failures. A single node failure can cascade into system-wide downtime, highlighting the critical need for resilient fault tolerance mechanisms to maintain training efficiency.

AI accelerators and network devices frequently encounter errors, including hardware failures, software bugs, and communication timeouts. Critically, failure frequency exhibits a strong positive correlation with cluster size.
For instance, training of the Bloom model experiences 1-2 GPU failures per week on a cluster of 384 GPUs \cite{le2023bloom}, while the OPT-175B model encountered approximately 110 failures over a two-month period on a cluster of 992 GPUs \cite{msft2022metaseq}. Additionally,  pretraining of Meta's LLaMA3 experienced 466 job interruptions during a 54-day period on a cluster of 16, 384 GPUs \cite{dubey2024llama}. As AI clusters grow to accommodate larger models, the probability of failures increases proportionally. This trend underscores the imperative for robust error detection and recovery mechanisms to maintain training continuity and operational efficiency.

The most conventional approach to anomaly recovery systems is periodic checkpointing, wherein model states are saved to persistent storage at fixed intervals and restored after failures.
However, this method suffers from two critical limitations:
(1) The substantial I/O overhead incurred during checkpoint saving and loading grows proportionally with model size, creating a significant bottleneck as modern LLMs scale to hundreds of gigabytes or terabytes. This I/O burden dramatically slows overall training progress;
(2) When a failure occurs, recovery requires all clusters to revert to the last checkpoint, discarding all intermediate computations since the last checkpoint. Statistically, this results in approximately half of the work between checkpoints being redundantly recomputed per failure.
These limitations create an inherent tension in system design: increasing checkpoint frequency reduces recovery time but exacerbate I/O overhead, while decreasing frequency minimizes checkpointing costs at the expense of longer recovery periods.
This fundamental trade-off poses a major obstacle to achieving reliable large-scale training, motivating the development of more efficient fault-tolerance paradigms for large-scale LLM training.

Recovery Point Objective (RPO) and Recovery Time Objective (RTO) are critical metrics in disaster recovery planning\cite{omar2013evaluating}. 
Both metrics are also suitable for evaluating the efficacy of fault recovery systems in LLM training.
RPO defines the maximum tolerable data loss, measured from the most recent backup to the point of system failure. 
In the context of LLM training, RPO corresponds to the potential loss of training progress between the latest checkpoint and the failure point.
RTO, on the other hand, refers to the maximum acceptable duration for restoring system after an disaster. 
For LLM training, RTO is determined by the time required to restart the training process following a failure.
While RPO aims to minimize data loss, RTO focuses on reducing downtime. 
An effective fault recovery system for LLM training must strike an optimal balance between these two objectives, ensuring efficient recovery with minimal disruption to training job.

In this paper, we propose FlashRecovery, a fast and low-cost LLM training fault recovery system. Its key features and innovations include:
\begin{itemize}
  \item Fast recovery with optimal RTO. Our system actively detects hardware and software failures within seconds, enabling immediate response. And the recovery time is nearly scalability-agnostic, ensuring consistent performance regardless of training cluster size.
  \item Low-cost recovery with optimal RPO. The system restricts computation lost due to a failure recovery to a single training step and eliminates the need for periodic checkpointing.
\end{itemize}

The  main contributions of this paper are as follows:
\begin{itemize}
  \item A theoretical quantification is done to analyze the recovery overhead.
  \item RPO and RTO are introduced to evaluate the performance of fault recovery systems.
  \item A fast and low-cost recovery solution is devised and implemented for large-scale LLMs training.
  \item Our solution, FlashRecovery, is validated on a large computing cluster with 10, 000 devices.
\end{itemize}

\section{Recovery Overhead Analysis} \label{sec:model}

\begin{figure*}[htbp]
  \centering
  \includegraphics[width=1\linewidth]{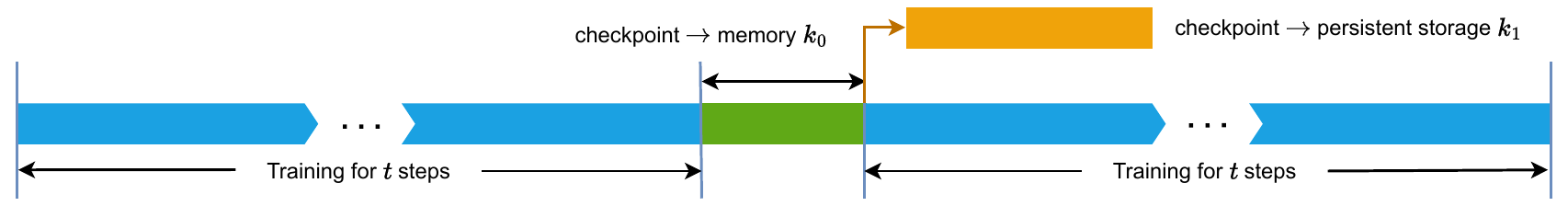}
  \caption{Alternative procedures (training and checkpointing) of normal training tasks.}
  \label{fig:proc:normal}
\end{figure*}

There are two procedures in common tasks of training deep learning models, which are alternatively executed by training processes (Fig. \ref{fig:proc:normal}):
\begin{enumerate}
  \item Training. Each training process does forward, backward and optimizer step to fit weights of a model.
  \item Checkpointing. Model parameters, optimizer states and other necessary information are stashed periodically (every $t$ steps) in case of possible failures.
   Usually checkpoints can be dumped to hosts' memory (procedure $k_0$ in Fig \ref{fig:proc:normal}) and then to persistent storage (procedure $k_1$ in Fig \ref{fig:proc:normal}).
\end{enumerate}

\begin{figure*}[htbp]
  \centering
  \includegraphics[width=1\linewidth]{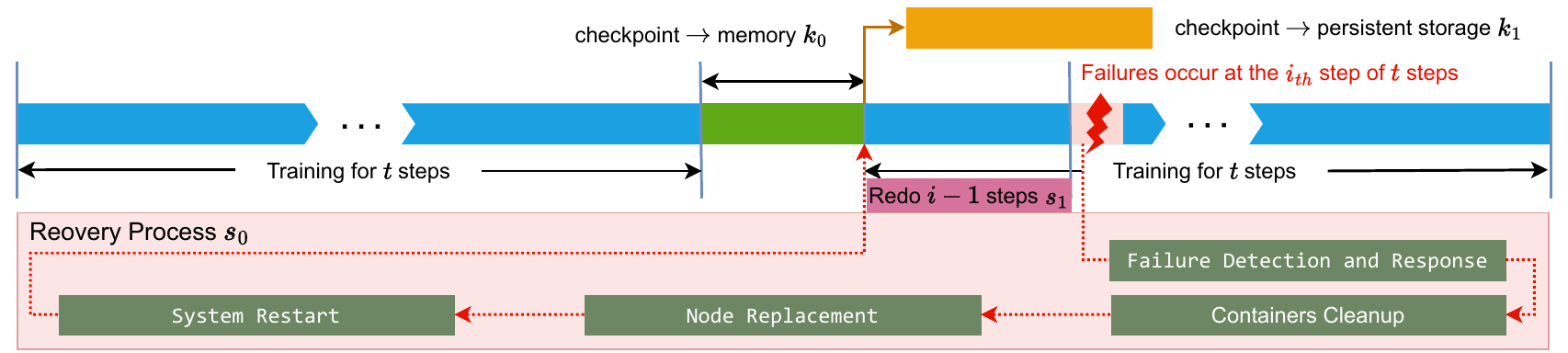}
  \caption{The standard recovery process from failures.}
  \label{fig:proc:failure}
\end{figure*}

\paragraph{Conventional Failure Recovery Protocol} When a failure occurs, a standard strategy for recovery typically involves the following steps (Fig. \ref{fig:proc:failure}): 
\begin{enumerate}
  \item Failure Detection and Response. The monitoring system identifies the failure and triggers recovery protocols. This phase inevitably incurs some latency between fault occurrence and system response due to detection overhead.
  \item Container\footnote{Modern clusters for deep learning generally use Docker to isolate physical environments and thus training processes run in containers.} Cleanup. The training process is halted and all containers on normal nodes are terminated.  
  \item Node Replacement. The faulty node is decommissioned and replaced with a new healthy one, which then rejoins the training cluster.
  \item System Restart. All containers are restarted across nodes, and the communication group is re-established.
  \item Traning Resumption. The system loads the latest checkpoint and resumes the training process.
\end{enumerate}
While this recovery procedure successfully restores the training state after failures, it introduces significant inefficiencies. 
The most notable drawback is the waste of computational work. 
Since the system must revert to the last saved checkpoint, all training progress between the latest checkpoint and the failure point is discarded.

\paragraph{Recovery Overhead Modeling} Ideally, users can adjust $t$ to change the frequency of checkpointing. 
A large $t$ results in a lower checkpointing frequency and minimize the training overhead. 
However, infrequent checkpointing may incur significant costs in the event of a failure, as all devices will need to redo a substantial amount of work from a long-ago checkpoint\cite{gupta2024just}. 
Thus,  a careful trade-off must be made between recovery costs and training overhead. 

To address this issue, we model the recovery process from failures and analyze the overhead of a recovery quantitatively, i.e., the elapsed time of recovery from failures. Suppose that:
\begin{itemize} 
  \item $d$. A fixed training time period.
  \item $t$. The number of steps between two consecutive checkpoints, referred to as the checkpointing interval. $\frac{d}{t}$ is the times of checkpointing during $d$.
  \item $m$. The number of failures occurring in the cluster during time period $d$.
  \item $s_0$. The recovery overhead, encompassing failure detection, failure response, container cleanup, node replacement, system restart and training resumption. This term is associated with the Recovery Time Objective (RTO).
  \item $s_1$. The recomputation cost, representing the training time overhead due to rollback. Under the assumption that failures occur uniformly at random, this cost can be approximated as $\frac{t}{2}$, corresponding to the Recovery Point Objective (RPO).
  \item $k_0$. The time taken to dump checkpoints from AI clusters to host memory (non-overlapping with other operations).
  \item $k_1$. The time taken to dump checkpoints from host memory to persistent storage (may overlap with training).
\end{itemize}
In the traditional periodic checkpointing approach, the parameters $m$, $s_0$, $k_0$ can be treated as constants, $k_1$ is negligible as it overlaps with training, while the checkpointing interval $t$ and the recomputation cost $s_1\approx\frac{t}{2}$ is tunable. The total failure recovery time $\mathcal{F}(t)$ can thus be expressed as a function of $t$:
\begin{equation}
  \mathcal{F}(t)=m*(s_0+s_1)+\frac{d}{t}*k_0=m*(s_0+\frac{t}{2})+\frac{d}{t}*k_0 \label{eq:f1}
\end{equation}
where $m*(s_0+\frac{t}{2})$ represents failure recovery costs and $\frac{d}{t}*k_0$ denotes checkpointing overhead. By optimizing $\mathcal{F}(t)$ with respect to $t$, we derive the optimal checkpointimg interval $t^*$ that minimizes total recovery time. This is obtained by solving:
\begin{equation}
  \mathcal{F}'(t)=\frac{m}{2}-\frac{d*k_0}{t^2}=0 \label{eq:f2}
\end{equation}
\begin{equation}
  t^* = \sqrt{\frac{2d*k_0}{m}} \label{eq:f3}
\end{equation}
The corresponding minimized recovery time is:
\begin{equation}
  \mathcal{F}_{min} = m*s_0 + \sqrt{2d*k_0*m} \label{eq:f4}
\end{equation}

The equation (\ref{eq:f3}) reveals two critical observations:
\begin{enumerate} 
  \item A higher failure rate (i.e., larger $m$) necessitates more frequent checkpointing (i.e., smaller $t$) to achieve the minimized recovery time $\mathcal{F}_{min}$.
  \item Conversely, a larger checkpointing overhead $k_0$ demands a larger checkpoint interval $t$ to achieve $\mathcal{F}_{min}$.
\end{enumerate}
And equation (\ref{eq:f4}) further identifies three primary directions for minimizing $\mathcal{F}_{min}$:
\begin{enumerate}
  \item Enhancing the stability of equipments, i.e., decrease $m$, which usually is hard to achieve and does not work when more devices are deployed. For example, when fault rate of a device is $0.001$, the probability that 100 devices work correctly is $(1-0.001)^{100}=0.90479$. While when we decrease the fault rate of a device to $0.0001$ (a tenth of $0.001$), the probability that 1000 devices work correctly is $(1-0.0001)^{1000}=0.90483$. It can be concluded that the improvements on stability of devices are mostly liked to be canceled out due to a larger scale of devices.
  \item Decreasing recovery overhead ($s_0$). $s_0$ typically grows with cluster size due to distributed coordination overhead. Decoupling $s_0$ from cluster scale, thus making $s_0$ become a cluster-size-agnostic constant, is a possible optimization goal.
  \item Reducing checkpoint overhead ($k_0$). A variety of checkpoint performance optimization methods have been developed to minimize $k_0$. However,  checkpoint-free recovery mechanism could achieve $k_0=0$, eliminating checkpointing overhead entirely. Furthermore, checkpoint-free approach naturally separates the  recomputation cost ($s_1$) from the checkpoint interval($t$), as the latter is not exist in a checkpoint-free system. 
\end{enumerate}

\begin{figure*}[htbp]
  \centering
  \includegraphics[width=1\linewidth]{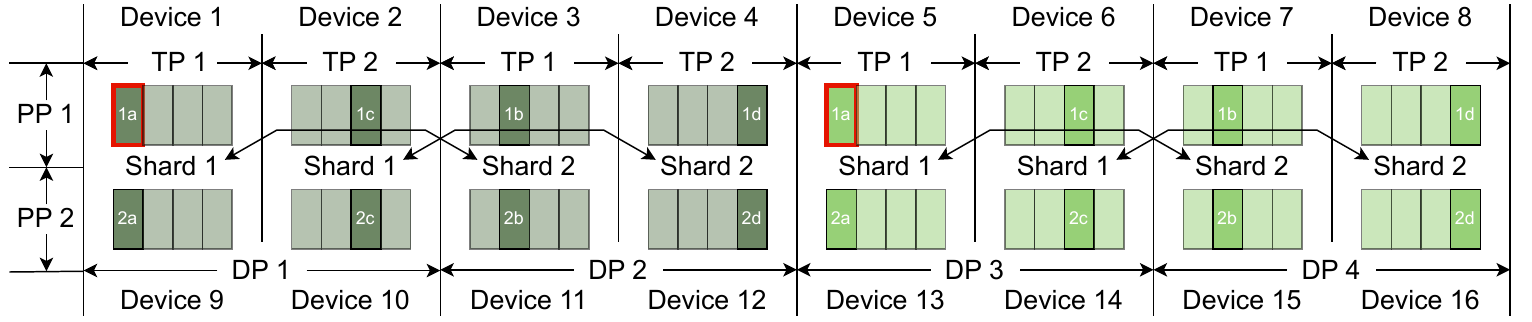}
  \caption{A combination of data parallelism, tensor parallelism, pipeline parallelism and ZeRO / FSDP parallelism. A frame represents a shard of parameters and the ones with the same id denotes the replicated parameters (e.g., the red frames are replicas of each other).} \label{fig:parallel}
  \vspace{-5pt}
\end{figure*}

In the following section, we elaborate on the key ideas of our recovery mechanism and its implementation. The quantitative analysis of our recovery mechanism and the system's limitations are also presented.

\section{Methodology}

\subsection{Motivation} \label{sec:motivation}

The scale of training datasets and the size of models in the state-of-the-art LLMs have grown at a exponential rate \cite{narayanan2021efficient}. There is no chance to fit the parameters of LLMs in the main memory of even the largest AI accelerator \cite{huang2019gpipe,rajbhandari2020zero,narayanan2021efficient,korthikanti2023reducing}. And vast amounts of data also require a lot of operations leading to a unbearably long training time on single device. Thus various model parallelism techniques have been proposed to address these challenges, such as:
\begin{itemize}
  \item Data parallelism (DP). Each process has a copy of the full model and aggregates the gradients periodically to ensure the consistency between model copies.
  \item Pipeline parallelism (PP). The layers of a model are sharded across multiple devices. A batch is split into multiple micro-batches, which are then pipelined across different pipeline stages. There are several possible ways of scheduling forward and backward of micro-batches, e.g., PipeDream-1F1B \cite{huang2019gpipe}, 1F1B-interleave \cite{narayanan2021efficient}, etc.
  \item Tensor parallelism (TP). Tensors in the two-layer multi-layer perceptron (MLP) and the attention module can be split along columns, rows or heads and matrix multiplications (GEMM) are then performed in these partitioned tensors, which reduces the memory footprint \cite{narayanan2021efficient}.
  \item Zero redundancy Optimizer (ZeRO) / Fully Shared Data Parallel (FSDP). Unlike basic data parallelism where model states are replicated across processes, ZeRO / FSDP partitions model states in stead, to scale the model size linearly with the number of devices \cite{rajbhandari2020zero}. ZeRO / FSDP is a special data parallelism with sharded model states.
\end{itemize}
A combination of different parallelism techniques can be deployed simultaneously, which is demonstrated in Fig \ref{fig:parallel}.

To process datasets more efficiently, data parallelism is often deployed with other parallelism techniques. With the deployment of data parallelism, we can confirm that there must be at least one replica of a model state for each process. Suppose the degree of data parallelism is $N$, then the number of replicas of a model state on single device is $N-1$. Given a fault rate $0.001$ of a device and $N=4$, the probability that none of those $N$ devices in a data parallelism group work correctly is $0.001^N=1e^{-12}$, which is extremely small and means that it is not likely to lose all of copies of a model state. Based on the 
quantitative analysis, we can conclude that it is a robust way to recovery from failures based on the replicas in a data parallelism group. The bigger the degree of data parallelism is, the more robust the recovery mechanism is.

\vspace{-3pt}
\subsection{Overview of the System Architecture}

In response to the challenges mentioned above, we propose \textit{FlashRecovery} ---- a fast and low-cost recovery mechanism comprising three key modules:

\begin{enumerate}
  \item Active and real-time failure detection. 
  This module employs a heartbeat mechanism coupled with device plugins to continuously monitor node performance and operational status.
  It detects failures within seconds and immediately broadcasts system-wide notifications. 
  Compared to conventional timeout-based approaches, our method significantly reduces monitoring overhead.
  \item Scale-independent task restart. 
  This module handles failures and restarts training jobs. 
  Unlike conventional approaches that indiscriminately terminate and restart all training processes, our system localizes the impact of failures by selectively substituting the faulty node with a healthy one. By limiting the number of nodes requiring restart, our approach decouples task restart time from cluster scale. We further accelerate recovery process through parallelized TCP Store initialization and by eliminating ranktable negotiations between devices, ensuring communication group setup remains independent of cluster size.
  \item Checkpoint-free recovery within one step. 
  This module takes a replica of the model state from other normal devices in the data parallelism group to restore failed processes.
  By leveraging data parallel redundancy, it eliminates checkpointing requirements while guaranteeing that at most one training step's progress is lost. 
\end{enumerate}

The details of design and implementation of our system are described in the subsequent sections.

\begin{figure}[htbp]
    \centering
    \includegraphics[width=1.\linewidth]{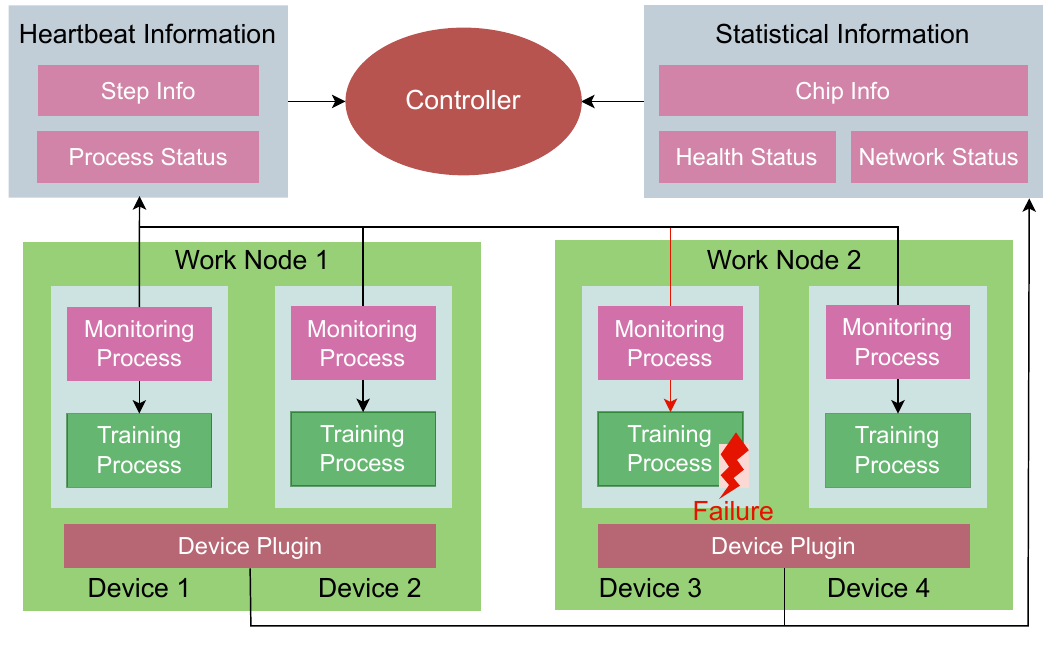}
    \caption{The architecture of the system and the workflow of the failure detection.}\label{fig:module:detect}
    \vspace{-7pt}
\end{figure}

\subsection{Active Real-time Failure Detection}

In the absence of an additional failure detection module, training processes traditionally identify occurrences of failures from other devices by a hang during collective communication, which can last up to 30 minutes in PyTorch. To address the inefficiency of the passive manner of sensing failures, we propose a novel failure detection mechanism which detects failures actively and achieve fast failure identification. The workflow of the mechanism is illustrated in Fig. \ref{fig:module:detect} and it consists of three components:

\begin{enumerate}
  \item Controller. A controller is a global service, which collects failure report from device plugins and monitoring processes. It also decides strategies to handle failures that occur at different stages and reschedules all processes after failures.
  \item Monitoring Processes. Monitoring Processes are created and run with every training process. They are able to monitor the health status of the associated training process and collect other necessary information (e.g. current step number) for recovery, which are reported to the controller periodically.
  \item Device Plugins. Device plugin is a component that is installed on every node and is able to report various statistical information of devices on a node, including chip info, health and network status. The information helps to determine the status of a device.
\end{enumerate}
Both heartbeat mechanism and device plugins provide a active ability to detect failures, which helps to detect a failure in seconds. When a failure occurs and the faulty device is confirmed, the controller decides the recovery strategy for every nodes and reschedules all training processes. Subsequently, our system restarts and recoveries the training task, which is to be introduced in the next two sections.

\vspace{5pt}
\subsection{Scale-Independent Task Restart}\label{sec:restart}

Traditional system restart methods often exhibit a linear increase in time consumption as cluster scale grows, primarily due to three factors:
\begin{itemize}
  \item Inefficient Container Management. These approaches destroy and recreate all containers indiscriminately, even when the training environment remains intact on normal nodes. This forces the system to wait for the slowest container initialization, creating a performance  bottleneck. Since container startup times follow a normal distribution, larger clusters inevitably encounter longer tail latencies, leading to linear growth in total reconstruction time.
  \item Scale-dependent Communication Group Establishment. The restart of all containers is accompanied by the creation of a new global communication group. This procedure entails the establishment of multiple communication links, followed by the execution of data exchange operations across these newly formed connections. 
  Notably, as the cluster size increases, both the volume of data exchanged and the number of required communication links grow proportionally. In the unoptimized implementation, these tasks are executed serially within a single process, leading to a linear increase in time complexity relative to the number of nodes.
  \item I/O Overhead During Training Process Initialization. 
  The training process initialization phase requires loading both the python environment (which may consist of tens of thousands of small files) and checkpoint (which can scale up to hundreds of gigabytes or terabytes). When thousands of containers restart simultaneously, massive parallel access to shared storage resources leads to severe I/O pressure. This bottleneck severely degrades initialization performance and further increases reconstruction latency.
\end{itemize}
These factors collectively constrain the scalability of traditional recovery mechanisms, motivating the need for a more efficient approach.

In FlashRecovery, as illustrated in Fig \ref{fig:module:restart}, normal nodes, abnormal nodes, and the controller exhibit distinct behavioral patterns during the task restart process. We optimize the restart process, which can be systematically divided into three stages:
\begin{enumerate}
    \item Node Rescheduling with Limited Recreation. Upon detecting a fault, the controller initiates a concurrent recovery protocol for different nodes.
    First, it dispatches termination signals to every normal nodes, instructing them to suspend their training processes. These nodes then transition to a standby state, awaiting for another continue signal from the controller to restart their training processes.
    Simultaneously, the controller executes node rescheduling, substituting faulty nodes with healthy ones. The newly allocated nodes execute the training script, initialize communication, and notify the controller to update the ranktable accordingly. The restart procedure for new joined nodes and the suspension of training on normal nodes are executed concurrently. Through applying different strategies for normal and faulty nodes respectively, we restrict the number of recreated nodes to only those encountering errors and reduce the unnecessary container recreation, which make the restart process independent of the scale of the training cluster and faster.
    \begin{figure}[h]
        \vspace{-7pt}
        \centering
        \includegraphics[width=\columnwidth]{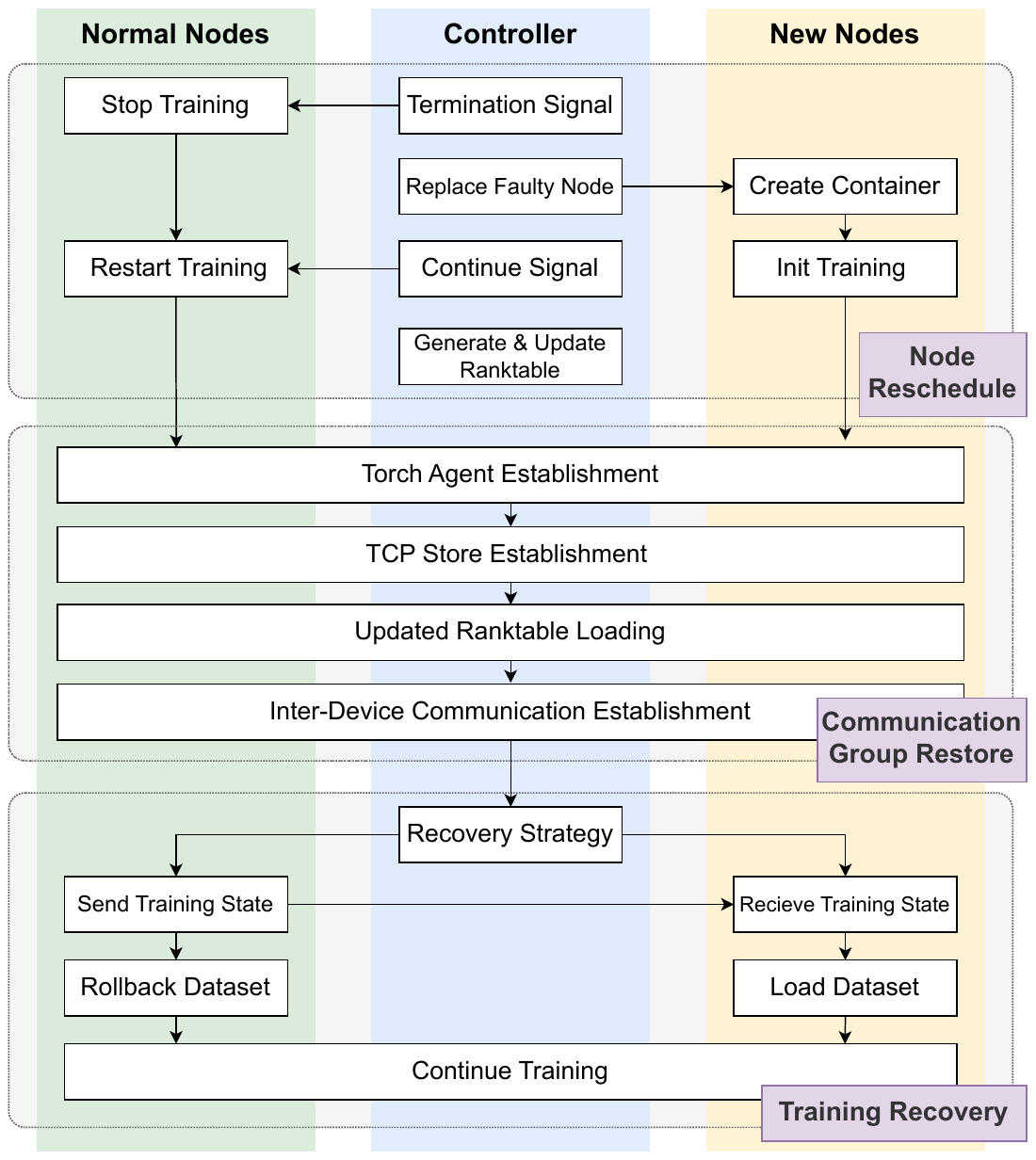}
        \caption{The task restart process in FlashRecovery.}
        \label{fig:module:restart}
        \vspace{-3pt}
    \end{figure}
    \item Optimized Communication Group Establishment. 
    The communication group establishment process is a critical step in the recovery process, as it involves the establishment of multiple communication links and data exchange operations across devices.
    Some procedures in this process are typically executed serially within a single process, leading to a time complexity relative to the number of nodes, i.e., $O(n)$.
    The process of establishing communication group usually can be decomposed into four procedures: 
    \begin{itemize}
        \item Torch Agent establishment. This procedure involves communication initialization and making connection with master node, which usually exhibits a relatively fixed time consumption. 
        \item TCP Store establishment. In contrast to Torch Agent establishment, TCP Store generally establishes in a serialized way, resulting in a linear time consumption dependent on cluster size. To improve efficiency, we apply a optimized strategy and parallelize the establishment of TCP Store, which reduces the time complexity from $O(n)$ to $O(\frac{n}{p})$ ($n$ is the scale of a cluster and $p$ is the degree of parallelization). 
        \item Updated ranktable loading. The ranktable records the resource information of the entire cluster for inter-device communication establishment. Originally, master node collects information from every node and then generates a global ranktable, which is sent to every node later. The generation and distribution of ranktable is executed serially and thus the time complexity is $O(n)$. In contrast, the controller in our system maintains a global ranktable in a shared file across nodes. Every device is able to load the latest ranktable from the file directly without any collection and distribution of ranktable, reducing the time complexity to $O(1)$.
        \item Inter-device communication establishment. The inter-device link establishment also adopts highly parallelized measures and the time is primarily dependent on the number of communication neighbors of the communication operators rather than the cluster size.
    \end{itemize}
    \item Training State Recovery. The controller employs distinct recovery strategies for normal nodes and faulty nodes. It also determines the device on normal nodes whose model state will be used to restore the training on newly scheduled nodes. A comprehensive analysis of this approach will be presented in the following section. 
\end{enumerate}

\subsection{Checkpoint-Free Recovery within One Step}

In general, model states of a specific step can be recovered totally from the last checkpoint, which requires extra training from the step of the checkpoint to the step with failures. To reduce the overhead resulting from the redone training, we propose a checkpoint-free recovery approach as follows:

\begin{enumerate}
  \item Restoration. We use data parallelism replicas to restore consistent model states of the restarted processes rather than a checkpoint. In this way, training processes hold model states of the $i_{\mathrm{th}}$ or $(i+1)_{\mathrm{th}}$ step. The specific step is depended on the phase where a failure occurs. We will describe the details of model state restoration and step determination in the following.
  \item Rollback. The iterator of dataset is rolled back to the step aligned with the model state ($i_{\mathrm{th}}$ or $(i+1)_{\mathrm{th}}$ step).
  \item Continue Training. The training process is totally recovered after restoration and rollback, and the training loop continues for a new batch.
\end{enumerate}

\begin{figure}[htbp]
  \begin{subfigure}[b]{0.24\textwidth}
    \centering
    \includegraphics[width=1.\linewidth]{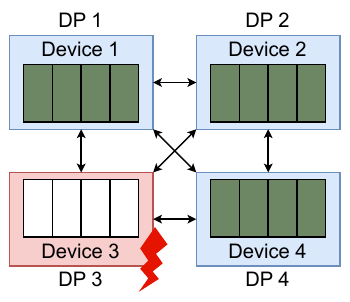}
    \caption{Vanilla DP restoration.}
  \end{subfigure}
  \begin{subfigure}[b]{0.24\textwidth}
    \centering
    \includegraphics[width=1.02\linewidth]{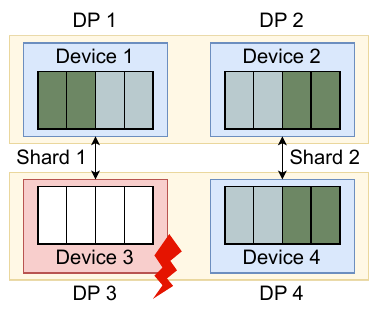}
    \caption{ZeRO/FSDP restoration.}
\end{subfigure}
\caption{Model states restoration with different parallelism.}
\label{fig:restore}
\vspace{-3pt}
\end{figure}

\paragraph{Model states Restoration} As described in \ref{sec:motivation}, we can restore the model states of recreated processes from data parallelism replicas. In this case, any checkpoints is not required and model states in recreated processes can be restored with collective communication. Because the controller has the global failure information, we can easily determine if we have a model state replica for the faulty process. Actually, if there is at least one training process of the same DP group on the normal node, we can restore the model states of other processes on the faulty node. We support model states restoration from two kinds of data parallelism: (1) vanilla data parallelism, and (2) data parallelism with ZeRO / FSDP. The model states restoration for these two parallelisms are illustrated in Fig. \ref{fig:restore}. Furthermore, our system also supports model states restoration from a combination of various data and model parallelism, which is demonstrated in Fig. \ref{fig:parallel}.

\paragraph{Step $i$ or step $i+1$ to resume?} Suppose that a failure occurs at the $i_{\mathrm{th}}$ step, the phase where a failure occurs determines where to resume the training, which can be divided into two cases:
\begin{itemize}
    \item Failures occur during forward and backward. In this case, the parameters of the model have not been updated for the next step and the training process can be resumed from the $i_{\mathrm{th}}$ step.
  \item Failures occur during optimizer step. In this case, despite it is difficult to determine which parameters on a device have been updated, it can be confirmed that the parameters of a normal device will be updated. Therefore, the parameters from recreated processes is able to be restored from those updated parameters on normal devices. Then the training should be resumed from the $(i+1)_{\mathrm{th}}$ step.
\end{itemize}

\begin{figure}[htbp]
  \centering
  \includegraphics[width=1\linewidth]{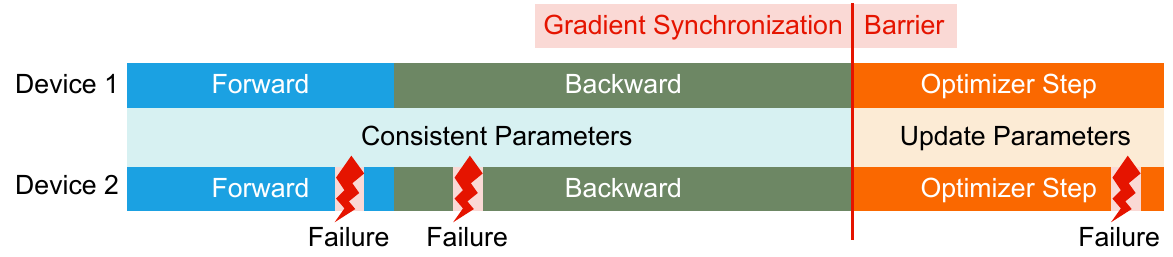}
  \caption{Training processes with data parallelism and barrier before optimizer step.}
  \label{fig:db:sync:barrier}
  \vspace{-3pt}
\end{figure}

Since devices execute asynchronously, we can not guarantee that all devices in the phase of forward and backward or the phase of optimizer step, which means above step determination strategy fails to be applied in practice. Fortunately, we can solve this problem with a easy barrier mechanism as demonstrated in Fig. \ref{fig:db:sync:barrier}. A barrier operation is added just in the beginning of the optimizer step, which achieves:

\begin{itemize}
  \item When a failure occurs during forward and backward, other normal processes must be hung before optimizer step due to the barrier operation, which ensures that all processes must be in the phase of forward and backward.
  \item When a failure occurs in a process during optimizer step, the barrier operation indicates that the process must have synchronized with other processes. In other words, all processes have entered the phase of optimizer step.
\end{itemize}
A synchronous barrier operation almost does not introduce any extra latency since we can merge the barrier operation and the last synchronization ---- gradient synchronization (by all-reduce).

\paragraph{The moment to stop, clean and reset} When we do restarting, the controller stops and cleans the kernels of the task-queue of every device and resets all devices on normal nodes, which destroys the running state of a process. When a failure occurs during forward and backward, it makes no sense to continue the execution of kernels because the model states of $i_{\mathrm{th}}$ step have not been updated. But when a failure occurs during optimizer step, the execution of kernels can be continued and we can restore the model states on the faulty node with the updated parameters (from the $(i+1)_{\mathrm{th}}$ step). We design a mechanism to determine when to issue "stop", "clean" and "reset" instructions from the controller to devices on normal nodes. The mechanism is implemented with step tags and includes the following steps:
\begin{enumerate}
  \item Set $\mathrm{step}=i$ for every training processes at the beginning of forward phase.
  \item The controller receives step tags from every device by the heartbeat mechanism.
  \item When a failure occurs during forward and backward, the controller receives $\mathrm{step}=i$ from all normal devices except those on the faulty node, we can confirm that all normal processes are in the phase of forward or backward and is able to issue "stop", "clean" and "reset" immediately without any side effect. 
  \item Set $\mathrm{step}=-1$ for every training processes at the beginning of optimizer step.
  \item Set $\mathrm{step}=i+1$ when a normal training process completes the optimizer step,  
  \item When a failure occurs during optimizer step, the controller receives $\mathrm{step}=i+1$ from all normal devices except those on faulty node, we can confirm the end of optimizer step of all processes on normal nodes. At this moment, the "stop", "clean" and "reset" instructions can be issued without any side effect and model states of training processes on the faulty node can be restored from the updated parameters at $\mathrm{step}=i+1$.
\end{enumerate}

\begin{figure*}[htbp]
  \centering
  \begin{subfigure}[b]{1\textwidth}
    \centering
    \includegraphics[width=1\linewidth]{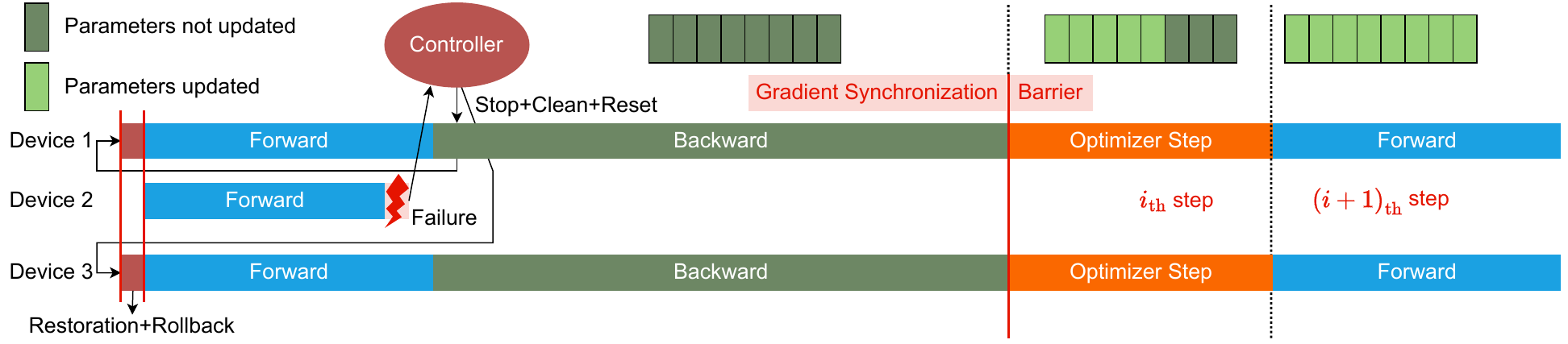}
    \caption{The recovery process from the $i_{\mathrm{th}}$ step when failures occur during forward and backward.}
    \label{fig:resume:a}
    \vspace{3pt}
  \end{subfigure}
  \begin{subfigure}[b]{1\textwidth}
    \centering
    \includegraphics[width=1\linewidth]{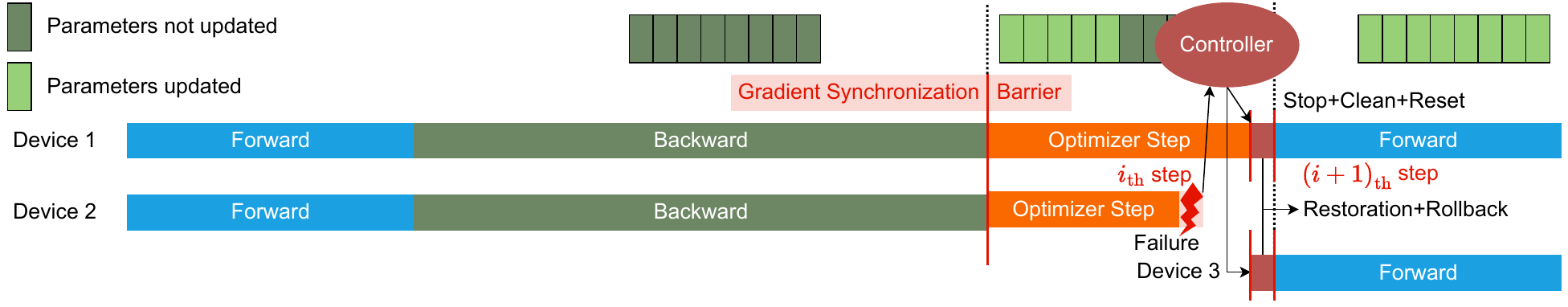}
    \caption{The recovery process from the $(i+1)_{\mathrm{th}}$ step when failures occur during optimizer step.}
    \label{fig:resume:b}
  \end{subfigure}
  \caption{The recovery processes for failures in different phases of a training step.}
  \label{fig:resume}
  \vspace{-10pt}
\end{figure*}

Finally, detailed recovery process of our system is illustrated in Fig. \ref{fig:resume}. The recovery process for failures in the phase of forward and backward is shown in Fig. \ref{fig:resume:a} and the recovery process for failures in the phase of optimizer step is demonstrated in Fig. \ref{fig:resume:b}.

\subsection{Recovery Overhead Analysis of Our Method}

Above all, our approach can recover training task with DP replicas and remove the extra redone training from checkpointing. 
The recovery time due to extra redone training ($s_1$) becomes a constant (approximately a bit longer than the running time of one training step). 
Moreover, the active failure detection mechanism can reduce the time of detecting a failure. 
And the scale-independent task restarting only restarts processes on faulty nodes, which is independent of the scale of a training task and further reduces the restarting overhead. Both of them reduce the recovery time $s_0$. 
Checkpoint-free recovery from DP replicas also removes the checkpointing overhead $\frac{d}{t}*k_0$. To sum up, the equation \ref{eq:f1} becomes:
\begin{equation}
  \mathcal{F} =m*(s_0'+s_1') \label{eq:f5}
\end{equation}
where $s'_0$ represents the optimized recovery overhead regardless of cluster size, and $s'_1$ denotes the optimized recomputation cost, which is limited to only one step.

\subsection{Limitations of Our Method}

Although our method substantially reduces recovery overhead, it still has some limitations:
\begin{enumerate}
  \item The method still cannot fully eliminate checkpointing in practice because there remains a small chance that all devices in the same DP group fails simultaneously, leaving no replica to restore the training processes on the faulty node. In such case, checkpointing overhead is still unavoidable. While the probability of this scenario is extremely low  (allowing reduced checkpointing frequency), our method still incurs significantly lower overhead than the vanilla approach.
  \item The solution requires tight coordination among devices, the training framework, and the global controller. This architectural complexity complicates integration and deployment. Future work should prioritize standardizing interfaces to improve usability.
  \item Current failure detection relies on active heartbeats and may not promptly identify processes stalled during computation or communication. Additionally, our failure categorization does not cover all possible error scenarios.
\end{enumerate}

\section{Experiments and Evaluations}

Our system is implemented on a computing cluster equipped with Huawei Ascend NPU and Kunpeng CPU\footnote{Ascend: \url{https://www.hiascend.com/}, Kunpeng: \url{https://www.hikunpeng.com/}}, which deploys more than 10, 000 NPUs.

\subsection{Failure Types And Ratios}

\begin{figure}[htbp]
    \centering
    \includegraphics[width=1.07\columnwidth]{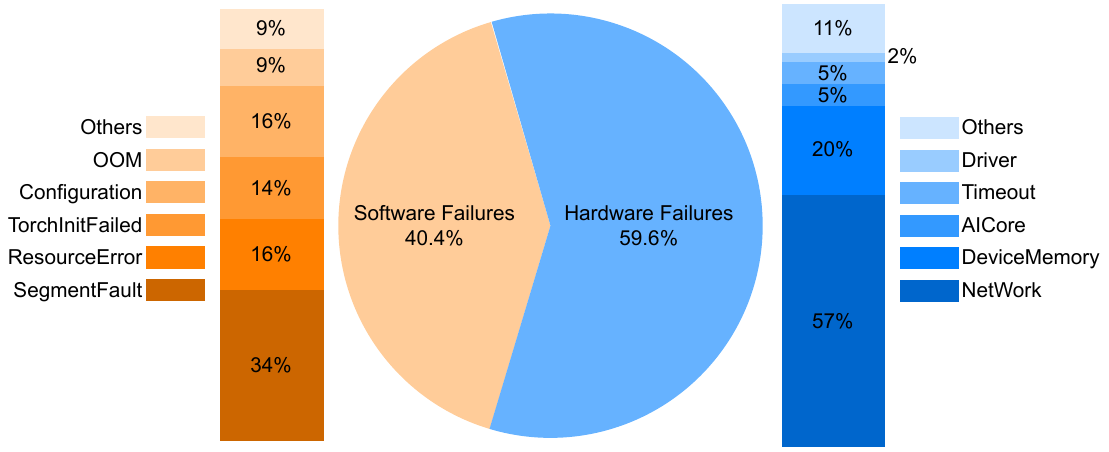}
    \caption{The types and frequencies of observed failures.}
    \label{fig:failures}
\end{figure}

We performed a comprehensive statistical analysis of failure occurrences in our training cluster Fig. \ref{fig:failures}.
The types of failures can be broadly categorized into two types: hardware failures and software failures. 
Our findings indicate that hardware failures constituted the majority, accounting for 59.6\% of all failures, while software failures represented 40.4\%.
Within hardware failures, network anomalies are the most prevalent, contributing to 57\% of cases, followed by Device Memory issues at 20\%. The remaining hardware failures included AICore failures, timeout errors, and driver-related problems. Additionally, 11\% of the observed failures could not be classified within our existing taxonomy.
Among software failures, segmentation faults are the most frequent, comprising 34\% of cases. Other notable software-related issues includes resource errors, torch initialization failed, configuration anomalies, and out-of-memory (OOM) errors. A small proportion (9\%) of software faults remains unclassified.

\subsection{Evaluation of Communication Group  Establishment}

This study implements a parallelized strategy for TCP Store establishment and an optimized ranktable update that eliminates any negotiation with the master node.

\begin{figure}[htbp]
    \centering
    \includegraphics[width=\columnwidth]{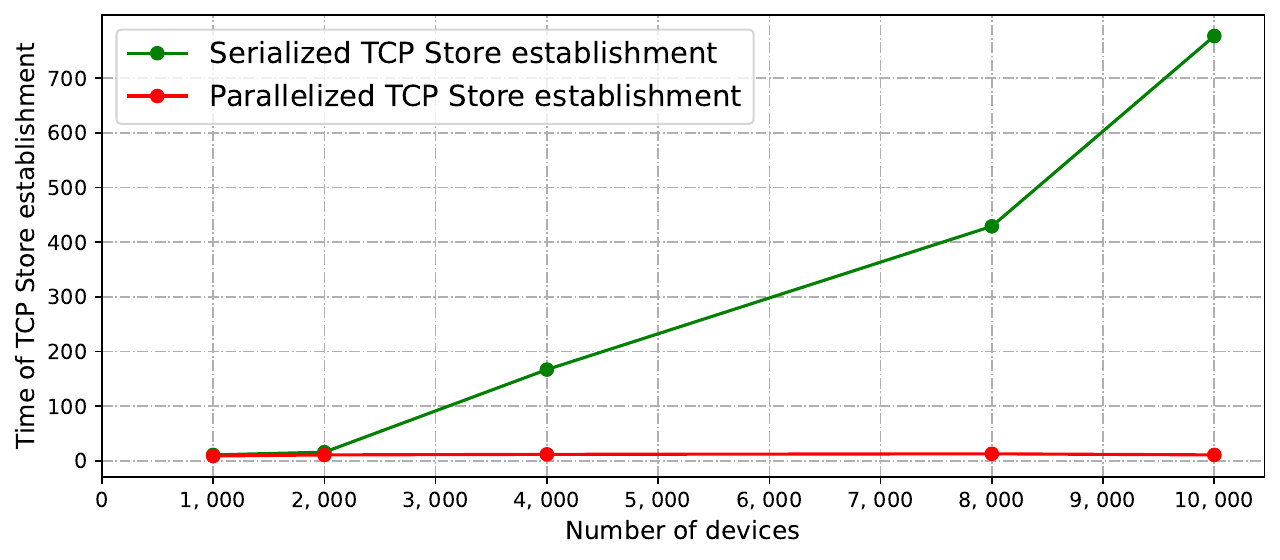}
    \caption{The time cost of TCP Store establishment under different approaches (in seconds).}
    \label{fig:tcp}
    \vspace{-10pt}
\end{figure}

First, we analyze the time required for TCP Store establishment under two approaches: (1) the serialized procedure, and (2) the parallelized procedure across varying scales of training clusters. Experimental results indicate that the original serialized link establishment method (green line in Fig. \ref{fig:tcp}) exhibits near-linear time complexity with respect to cluster scale. In contrast, our system uses the parallelized approach (red line in Fig. \ref{fig:tcp}) that significantly reduces the scaling coefficient and effectively decouples establishment overhead from cluster scale.

Secondly, we evaluate the time cost of ranktable updates before and after the introduction of a global ranktable. The default way for updating a ranktable requires message collection and distribution between every device and the master node, with a time cost proportional to the scale of the training cluster (i.e., $O(n)$).  This relationship is confirmed by observations in the first row of Tab. \ref{tab:recovery:ranktable}. In contrast, FlashRecovery’s controller maintains an up-to-date global ranktable, which can be loaded via a shared file. This eliminates communication overhead for ranktable updates and accelerates the process. Experimental results demonstrate that the time cost of updating the ranktable via a shared file  (bottom row of Tab. \ref{tab:recovery:ranktable}) is optimized to a constant (i.e., $O(1)$), significantly improving efficiency.

\begin{table}[htbp]
    \centering
    \caption{The time of ranktable updating (in seconds).}
    \label{tab:recovery:ranktable}
    \resizebox{\linewidth}{!}{
    \begin{tabular}{cccccc}
        \toprule
        Num. of Devices & 1, 000 & 4, 000 & 8, 000 & 16, 000 & 18, 000 \\
        \midrule
        Original ranktable updating & 8 & 31 & 60 & 176 & 249 \\
        Loading ranktable directly & 0.1 &  0.1 & $<0.5$ & $<0.5$ & $<0.5$ \\
        \bottomrule
    \end{tabular}
    }
    \vspace{-5pt}
\end{table}

\subsection{Assessment of Overall Failure Recovery Time}

We first evaluate the recovery time ($s_0$ in equation \ref{eq:f1}) of a vanilla method for different task scales, details of which are provided in Tab. \ref{tab:recovery:vanilla}. We use the default configuration of PyTorch. A failure is reported when a communication hang lasts 1, 800 seconds. The task restarting time increases linearly with the task scale, as illustrated in the last column of Tab. \ref{tab:recovery:vanilla}. Since a failure occur randomly between two consecutive checkpoints, we do not include the recomputation cost ($s_1$ in equation \ref{eq:f1}, i.e., the wasted work from the last checkpoint to the step that a failure occurs).

To assess the efficacy of FlashRecovery, we artificially inject faults during the standard training process and record the time taken for each stage in the recovery. From the 3rd column of Tab. \ref{tab:recovery:flashrecovery}, our system is able to sense a failure within 10 seconds. And from the 4th column of Tab. \ref{tab:recovery:flashrecovery}, the restart time in our system is almost invariable and becomes independent of the scale of a training cluster. In addition, according to the assumption about $s_1$, we use the average time of one step to estimate the time of redone training, i.e., half of the redone step (shown in the 6th column in Tab. \ref{tab:recovery:flashrecovery}). The variation on the scale of training clusters has much less impact on the total recovery time. Although we increase the number of devices from 32 to 4, 800, the total recovery time is still able to remains in 150 seconds, which increase about 52\% and is much less than the rising on the number of devices (150$\times$). In short, the results in Tab. \ref{tab:recovery:flashrecovery} show that the time required for failure recovery by our system is nearly scalability-agnostic.

\begin{table}[htbp]
  \centering
  \caption{The recovery time of vanilla recovery method at different task scales (in seconds).}
  \label{tab:recovery:vanilla}
  \resizebox{\linewidth}{!}{
    \begin{tabular}{cccc}
        \toprule
        Num. of Parameters & Num. of Devices & \makecell{Failure Detection Time
        } & \makecell{Task Restarting Time
        } \\
        \midrule
        175B & 1, 824 & 1, 800 & 231 \\
        175B & 3, 936 & 1, 800 & 801 \\
        175B & 5, 472 & 1, 800 & 1, 115 \\
        \bottomrule
    \end{tabular}
    }
    \vspace{-5pt}
\end{table}

\begin{table*}[htbp]
  \centering
  \caption{Recovery time of our method for different task scales and model sizes (in seconds).}
  \label{tab:recovery:flashrecovery}
  \resizebox{\linewidth}{!}{
  \begin{tabular}{ccccccc}
    \toprule
    \multirow{2}{*}{Num. of Parameters} & \multirow{2}{*}{Num. of Devices} & \multirow{2}{*}{\makecell{Failure Detection Time 
    }} & \multirow{2}{*}{\makecell{Task Restarting Time 
    }} & \multicolumn{2}{c}{\makecell{Redone Training 
    }} & \multirow{2}{*}{\makecell{Total Time 
    }} 
    \\
    \cline{5-6}
    & & & & step & step/2  & \\
    \midrule
    7B & 32 & 6 & 88 & 6 & 3 & 97 \\
    7B & 960 & 6 & 92 & 6 & 3 & 101 \\
    70B & 80 & 4 & 84 & 4 & 2 & 90 \\
    70B & 800 & 9 & 92 & 20 & 10 & 111 \\
    70B & 960 & 8 & 78 & 24 & 12 & 98 \\
    70B & 2, 880 & 11 & 90 & 39 & 19.5 & 120.5 \\
    175B & 2, 880 & 10 & 90 & 79 & 39.5 & 139.5 \\
    175B & 4, 800 & 7 & 116 & 49 & 24.5 & 147.5 \\
    \bottomrule
  \end{tabular}
  }
  \vspace{-10pt}
\end{table*}

\section{Related Work}

\subsection{Comprehensive Failure Recovery System} 

A variety of comprehensive failure recovery systems have been proposed, mainly encompassing failure detection, task rescheduling, and/or training restart mechanisms.
TRANSOM \cite{wu2023transom} is a fault-tolerant system for large-scale model training, which integrates an automatic fault tolerance and recovery mechanism, an automatic anomaly detection system and the asynchronous checkpoint saving technology.
Unicron \cite{he2023unicron} is a workload manager designed for self-healing in LLM training, demonstrating up to a 1.9$\times$ improvement in training efficiency. 
Its key features include in-band error detection for real-time error identification, a dynamic cost aware plan generation mechanism for optimal reconfiguration, and a transition strategy to reduce downtime during state changes. 
MegaScale \cite{jiang2024megascale} has developed a set of diagnostic tools designed to monitor system components and events. These tools focus on identifying root causes of issues and deriving effective techniques to achieve fault tolerance and mitigate stragglers. 
As a result, MegaScale improves Model FLOPs Utilization (MFU) by 1.34$\times$ compared to Megatron-LM. 
MoC-System \cite{cai2024moc} is a fault tolerance system specifically designed for the sparse Mixture-of-Experts model.
It incorporates 1) partial experts checkpointing mechanism, 2) fully sharded checkpointing strategies, and 3) a two-level checkpointing management method to enhance reliability and efficiency.
However, these systems are often customized for specific scenarios, making it difficult to achieve claimed performance in different training environments and configurations.
Diversity in hardware, network architecture, and workload characteristics may complicate the implementation of these fault recovery systems.

\subsection{Checkpointing Performance Optimization} 

Checkpointing is a common practice for failure recovery in LLM training.
Due to the substantial sizes of LLMs, a straightforward checkpointing solution that directly dumps model states to persistent storage incurs significant I/O overhead.
To address this challenge, a range of checkpoint performance optimization methods has been developed
to minimize the associated overheads.

\paragraph{Checkpionting Speed Up} 

Checkpointing typically consists of two phases\cite{duan2024efficient}: 
(1) snapshot phase ($k_0$ in Fig. \ref{fig:proc:failure}), 
(2) persist phase ($k_1$ in Fig. \ref{fig:proc:failure}). 
To address the time cost associated with checkpointing, several approaches have been developed to decouple these two phases, which allow training to stall for only a few seconds during the snapshot phase while asynchronously persisting snapshots using dedicated background CPU processes.
DataStates-LLM \cite{maurya2024datastates} introduces a lazy asynchronous multi-level approach that overlaps checkpoint I/O with the immutable phases of forward and backward passes during training and speeds up checkpointing by a factor of 3× to 4.2×. CheckFreq \cite{mohan2021checkfreq} features a resumable data iterator and a pipelined two-phase checkpointing mechanism, by which a consistent and low-cost checkpointing at the iteration level is achieved. 
LightCheck \cite{chen2023cost} implements a fine-grained checkpointing scheme and a persistent memory (PM) manager. 
The checkpointing scheme pipelines checkpointing with computation and communication in a layer-wise manner.
Additionally, the persistent memory is mapped into GPU virtual memory space and separates the storage of metadata and data of tensors in PM. 

Some other approaches aim to reduce I/O overhead during checkpointing.
Check-N-Run \cite{eisenman2022check} takes advantage of the fact that only a fraction of the recommendation model is updated in each iteration and proposes an incremental checkpointing strategy, which is primarily limited to recommendation models.
DeepFreeze \cite{nicolae2020deepfreeze} combines lightweight serialization, sharding checkpointing across data-parallel GPUs, and the augmentation of the execution graph to asynchronously mask the overhead of capturing weights from tensors.
Universal checkpointing \cite{lian2024universal} decouples distributed checkpoint storage from parallelism methods, providing the flexibility to resume operations using various parallelism strategies and hardware configurations.
ByteCheckpoint \cite{wan2024bytecheckpoint} advocates for a parallelism-agnostic checkpoint representation and a generic workflow for saving and loading checkpoints.

Although the methods mentioned above can significantly reduce periodic checkpointing time costs, 
the enlarged checkpointing duration cannot be fully overlapped with the training process, a trade-off remains between checkpointing frequency and the time lost during recovery from failures.

\paragraph{Just-In-Time Checkpointing} 

Several approaches aim to avoid checkpointing until an error occurs, which can reduce the steady-state overhead associated with periodic checkpointing while minimizing wasted work during recovery.
Just-In-Time Checkpointing \cite{gupta2024just} introduces user-level recovery solutions and transparent error recovery mechanisms.
It captures just-in-time checkpoints immediately after failures occur, allowing training to resume seamlessly from these JIT checkpoints.
The Update-Undo Mechanism, proposed by Swift \cite{zhong2023swift}, enables failure recovery with replicas of a model state in data parallelism.
Additionally, Swift advocates for employing asynchronous logging methods to expedite failure recovery in pipeline parallelism.
While both Just-In-Time Checkpointing and Swift offer improvements in reducing steady-state overhead, they still require checkpoints when failures occur, which can stall the recovery process and lead to delays in resuming training.
Parcae \cite{duan2024parcae} further eliminates the need to restart and roll back to previous checkpoints when failures occur. 
Instead, it transfers valid model states from the remaining healthy instances to reconfigured nodes. 
However, Parcae is specifically designed for LLM training on preemptible instances and requires adjustments to the parallelization strategy.

\section{Conclusions}

This paper presents the design and implementation of FlashRecovery, a fast and low-cost failure recovery system for large-scale training of LLMs. 
By incorporating a novel active and real-time failure detection mechanism, along with a scale-independent task restart mechanism and checkpoint-free recovery within one step, FlashRecovery can reduce the failure recovery time overhead of a training cluster with thousands of devices to less than 150 seconds and almost achieve the optimal RPO and RTO.
In the future, we will focus on standardizing the system's interfaces to improve usability, optimizing the recovery process and further reducing the total recovery time to 30 seconds for clusters with more than 10, 000 devices.

\bibliographystyle{IEEEtran}
\bibliography{IEEEabrv,references}

\end{document}